\documentclass[a4paper,12pt]{article}

\usepackage{amsmath,amssymb,amsfonts}
\usepackage[dvips]{graphicx}
\usepackage{epsfig}

\usepackage{color}
\usepackage{amsmath}
\usepackage{amsfonts}
\usepackage{verbatim}
\usepackage{amssymb}
\usepackage{graphicx}
\usepackage{amsmath,amssymb,calc}
\usepackage{bbm}
\usepackage{setspace}
\usepackage{color}
\usepackage{amsmath}
\usepackage{amsfonts}
\usepackage{verbatim}
\usepackage{amssymb}
\usepackage{graphicx}
\usepackage{array}
\usepackage{epstopdf}
\usepackage{amsmath,amssymb,amsfonts,graphicx}
\usepackage{epsfig}
\usepackage[left=2.4cm,top=3.3cm,right=2.4cm,bottom=3.3cm,bindingoffset=0cm]{geometry}

\newcommand{\beq}{\begin{eqnarray}}
\newcommand{\eeq}{\end{eqnarray}}
\newcommand{\bea}{\begin{eqnarray}}
\newcommand{\eea}{\end{eqnarray}}
\newcommand{\be}{\begin{equation}}
\newcommand{\ee}{\end{equation}}

\def\1{\mathbbm{1}}

\numberwithin{equation}{section}

\begin{document}

\title{
\begin{flushright}\ \vskip -1.5cm {\small {IFUP-TH-2016}}\end{flushright}
\vskip 20pt
\bf{ \Large Confinement, NonAbelian monopoles, and 2D ${\mathbb C}P^{N-1}$  model on the worldsheet of  finite-length strings}\footnote{Invited talk given at
CONF12, 'XII Quark Confinement and the Hadron Spectrum',  Thessaloniki,  August 29th - September the 3rd (2016)}
\vskip 20pt}\author{
Kenichi Konishi$^{(1,2)}$    \\[20pt]
{\em \small
$^{(1)}$Department of Physics ``E. Fermi", University of Pisa}\\[0pt]
{\em \small
Largo Pontecorvo, 3, Ed. C, 56127 Pisa, Italy}\\[3pt]
{\em \small
$^{(2)}$INFN, Sezione di Pisa,    
Largo Pontecorvo, 3, Ed. C, 56127 Pisa, Italy}  
\\
{ \footnotesize kenichi.konishi@unipi.it,}   
}
\date{November  2016}
\maketitle
\vskip 0pt

\begin{abstract}

Quark confinement is proposed to be   a  dual Meissner effect of nonAbelian kind.   Important hints come from
 physics of strongly-coupled infrared-fixed-point theories in ${\cal N}=2$  supersymmetric QCD, which turn into confining vacua under a small relevant perturbation. 
 The quest for the semiclassical origin of the nonAbelian monopoles, ubiquitous as the infrared degrees of freedom  in   supersymmetric 
 gauge theories, motivates us to study the quantum dynamics of 2D ${\mathbb C}P^{N-1}$  model defined on a finite-width worldstrip, with various boundary conditions. The model is found to possess a unique phase ("confinement phase"), independent of the length of the string, showing the quantum persistence of the nonAbelian monopole.

\end{abstract}

\newpage 

\section{Confinement as non-Abelian dual Meissner effect}
\label{intro}
\begin{figure}[h]
\centering
\includegraphics[width=12
cm,clip]{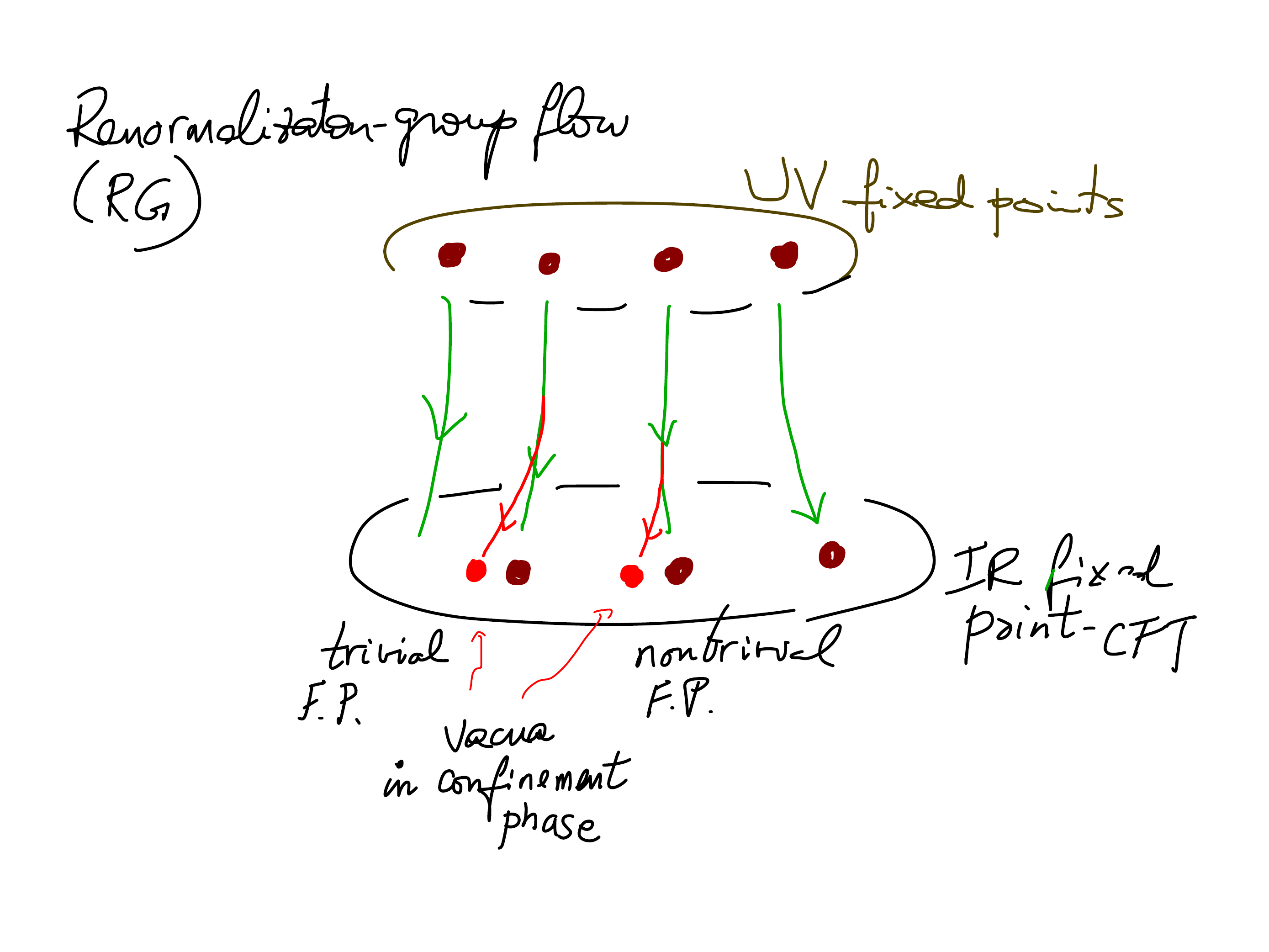}
\caption{RG flow of general aymptotically free gauge theories}
\label{fig-1}       % Give a unique label
\end{figure}

 In general asymptotic-free  gauge theories the renormalization-group (RG) flow looks like in Fig. \ref{fig-1}. 
The system in a UV fixed point,  which is a free Yang-Mills theory,  flows into an infrared fixed point (FP). In some cases the system approaches in the infrared a  FP conformally invariant theory (CFT), but before reaching it,  the RG path deviates from the one to CFT  to a nearby vacuum in a confinement phase, with mass generation and with the elementary charges (the quarks) confined by a vortex like chromoelectric strings.  Such a deviation in the RG flow at the final stage  is caused by some relevant operator, which could either be present in the original UV Lagrangian, or may be generated dynamically by the gauge dynamics itself.

One usually thinks that confinement and conformal invariance are two opposite notions:  in one case one has dynamical mass generation
and symmetry breaking, in the other, no condensation, no mass gap, no symmetry breaking.  How can such two  different phenomena be "close to each other"? 

The  answer is that in asymptotically-free gauge theories, the degrees of freedom in the low-energy effective theory can be different from those in the underlying UV theory. Instead of the original quarks and gluons,  the solitonic, magnetic monopoles and dual gauge fields may take up the role of  the carrier of the low-energy dynamics.  These degrees of freedom are {\it shared} between the infrared 
fixed-point theory and the nearby confining vacuum. 
There are many examples of  this kind of RG flows,  especially in the context 
of softly broken ${\cal N}=2 $ supersymmetric gauge theories \cite{SW1}-\cite{BGK}. 

In  theories described by such a RG flow,  {\it confinement is described by the same degrees of freedom characterizing the nearby CFT}.  
Thus the particular types of confining vacua  - the nature of the degrees of freedom and their interactions - are related to  the possible types of the nearby CFT. 
Such a correspondence could be particularly welcome, when the nearby fixed point is a strongly-coupled conformal theory, where relatively nonlocal degrees of freedom such as (nonAbelian) monopoles, dyons and quarks all appear simultaneously as  the infrared degrees of freedom. These theories are not described by a local Lagrangian and in general  are quite hard  to analyze \cite{SCFT},\cite{Eguchi}. Strong constraints on CFT may however allow us to draw  useful conclusions about the behavior of the system under the deformation into the nearby confinement vacuum, hence about the confinement mechanism itself. 

\section{QCD}
\label{sec-1}

Are these considerations relevant for the standard QCD?    The well-known idea, by Nambu, Mandelstam and  't Hooft \cite{tHooft},  is  that somehow the $SU(3)$ color gauge interactions  undergo  dynamical Abelianization
$   SU(3)   \to    U(1) \times U(1)  $\;, 
and  that   the low-energy degrees of freedom  (the monopoles of the two $U(1)$ (dual) theories)  somehow condense, and give rise to confinement in a scenario similar to  
the type II superconductivity.  The quarks (the chromoelectric charges) would  thus be  confined by a  
dual Meissner mechanism.

The problem is that  the presence of two confining strings  $\Pi_1(U(1) \times U(1)  )= {\mathbb Z}  \times  {\mathbb Z}\; 
$   implies the doubling of meson spectra,
not observed in Nature. Another difficulty is the following. If the confinement and chiral symmetry breaking are related,  the most natural  explanation is that the Abelian monopoles
carry the $SU_L(2)\times SU_R(2)$  chiral  symmetry  charges,  
$   M_i^j\;,  $
and their "diagonal" condensation 
\be    \langle M_i^j \rangle =    \delta_i^l\, \Lambda\;,   
\ee
explains the two phenomena at the same time.   However a $U(1)^2$ theory with  $N_f \times  N_f=4 $ scalar monopoles would have an accidental  $SU(4)$ symmetry, broken
to $SU(2)$,  implying  $12$ Nambu-Goldstone bosons,  again  not observed in Nature. Finally, $U(1)$ theories are weakly coupled at low energies: why should monopoles condense?

A possible alternative  is that the $SU(3)$ gauge symmetry dynamically reduces  as 
$   SU(3)   \to    SU(2) \times U(1)  \;;
$
the condensation of the nonAbelian monopoles would break the remaining gauge symmetries and produce a vacuum 
in which the quarks are confined in a dual Meissner effect, of a nonAbelian type. 
 As 
$  \Pi_1( SU(2) \times U(1)  )= {\mathbb Z}\;,  
$
the problem of the  meson spectrum doubling  is naturally solved.   A strongly interacting, nonAbelian magnetic monopoles may also  be welcome as  it could provide a natural dynamical mechanism for 
monopole condensation. 

One is aware of the fact that  a system of  strongly interacting nonAbelian monopoles can be something rather difficult to analyze.  Let us note however  that, unlike the "difficulties" mentioned earlier, 
this is a  difficulty for us, not for Nature.

\subsection{NonAbelian monopoles}

The concept of nonAbelian monopoles turned out to be peculiarly evasive. A famous difficulty is the so-called  topological  obstruction \cite{CDyons}\footnote{ It states that the "unbroken"  subgroup, e.g.,  $SU(N)$  in, 
$  SU(N+1) \to  SU(N) \times U(1)   \; 
$
(under which the 't Hooft-Polyakov monopoles of various broken subgroups $SU(2)$'s   are supposed to transform to each other),  cannot be defined globally, in the presence of a
monopole background. }   Another is an infinitesimal version of this: 
 gauge zeromodes around a soliton monopole solution are non-normalizable  \cite{DFHK}.  This means that the standard quantization procedure  to construct the  ${\underline N}$  gauge multiplet of monopoles through the action of the zero-mode creation operators
 does not work.  
 
Both these classical "difficulties" however miss the fundamental aspect  of the nonAbelian monopoles, expressed by the well-known GNO (Goddard-Nuyts-Olive) quantization condition \cite{GNO}.
 The asymptotic behavior of the monopole solution generated by the gauge symmetry breaking, $ G \to H$, can be written in an appropriate gauge as,    
\be        F_{ij} =  \epsilon_{ijk} \frac{r^k}{r^3}   (\beta \cdot  {\bf  H})\;,
\ee
where $\bf H$ is the Cartan subalgebra generators of the "unbroken" gauge group $H$,    and  $\beta$ is some  vector, characterizing each solution, having the number of components equal to the rank of the subgroup $H$.   GNO found that the consistency requires that 
\be     2 \beta \cdot  \alpha  \in {\mathbb Z}\;,  \label{GNO}
 \ee
where  $\alpha$ are the root vectors of $H$. Eq.~(\ref{GNO}) naturally generalizes  the famous Dirac quantization condition.   The GNO condition  states that the magnetic monopoles in the  broken gauge theory $G/H$ are labeled and classified by the weight vectors of the dual gauge group ${\tilde H}$, 
which is generated by the (nonzero) dual root vectors 
 \be   \alpha^* =   \frac{\alpha}{\alpha \cdot \alpha}\;,
 \ee
and not by the original group $H$ itself. 
Examples of pairs of dual groups are given in Table \ref{table1}.  

\begin{table}[h]
\centering
\caption{Examples of dual pairs of groups}
\label{table1}       % Give a unique label
% For LaTeX tables you can use
\begin{tabular}{ |c|c| }
\hline
  $H$  &  ${\tilde H}$       \\\hline
$U(N)$   & $U(N)$      \\   
$SU(N)$  & $SU(N)/Z_N$     \\   
$SO(N)$  & $Spin(N)$    \\
$SO(2N+1)$   &  $USp(2N)$    \\    \hline
\end{tabular}
% Or use
%\vspace*{5cm}  % with the correct table height
\end{table}

\section{Hints from ${\cal N}=2$ supersymmetric QCD}
\label{sec-2}

In spite of the classical "difficulties"  mentioned above, monopoles with nonAbelian (dual) gauge charges are ubiquitous in ${\cal N}=2$ supersymmetric gauge theories: they appear  
as low-energy effective degrees of freedom, in certain singular vacua \footnote{In simple singularities of the quantum moduli space of vacua the gauge system dynamically Abelianizes, i.e. reduces to the product of the $U(1)$ groups;  the monopoles appearing in such vacua are naturally Abelian.  Either by some particular choice of the bare quark mass parameters and/or in some points of quantum moduli space,  these Abelian singularities coalesce.  The monopoles appearing in these vacua are nonAbelian.  
}. They are pointlike particles of the low-energy effective theory, and act as carriers of the dual (magnetic) gauge charges, as well as of  the flavor symmetries of the underlying theory.
Under certain circumstances (e.g., appropriate ${\cal  N}=1$  relevant perturbation)  they condense and induce confinement and chiral symmetry breaking.  Exactly how this occurs depends on the
model considered, on the gauge and flavor groups, and on the particular infrared vacuum considered.  
\begin{description}
  \item[(i)]   Abelian dual superconductivity.  
  
  This occurs in ${\cal N}=2$  $SU(2)$ gauge theories \cite{SW1}, with number of flavors, $N_f=1,2,3$. The system dynamically Abelianizes, and Abelian monopoles condense upon some 
  ${\cal N}=1$ perturbation (the adjoint scalar mass term, which is a relevant operator).   Similar phenomena  occur in pure ${\cal N}=2$  $SU(N)$ theories in simple singularities of the vacuum moduli space \cite{SUN}. 
  This is a beautiful phenomenon, but appears to share little with the real-world QCD. 
  
  \item[(ii)]  NonAbelian monopole condensation.
  
  In ${\cal N}=2$   supersymmetric SQCD,   there are so-called $r$-vacua \cite{APS,CKM}, in which  the low-energy dual gauge group is 
  $SU(r)\times U(1)^{N - r}$.  The effective degrees of freedom there  turn out to be nonAbelian magnetic monopoles in  the  ${\underline r}$ representation of the dual $SU(r)$  gauge group, and   
  in ${\underline {N_f}}$ of the flavor $SU(N_f)$ symmetry group. The possible values of $r$  run up to   $r \le  [N_f/2]$.    The nonAbelian monopoles are weakly coupled, and 
  the  dual theory is infrared free.  Again,  these are interesting theories, but  do not resemble  the real-world strong interactions (weakly-coupled magnetic monopoles are not
  what we observe at low-energies).
    
  \item[(iii)]  NonAbelian monopoles interacting strongly. 
  
    The most intriguing systems  are the confining vacua arising from (or near) the strongly-coupled  infrared (super) conformal  points. 
  In general these systems involve relatively nonlocal degrees of freedom  (monopole, dyons and quarks, all becoming massless and interacting together), 
  which defy a local Lagrangian description.  Not surprisingly these are the least studied systems, but thanks to some key inputs by  Argyres, Seiberg and  others \cite{Argyres:2007cn}-\cite{Simone}
  there have been  some significant development  recently in understanding confinement near the so-called superconformal points of highest criticality (the Eguchi-Hori-Ito-Yang vacua) \cite{Eguchi}.

   \end{description}

\subsection{${\cal N}=1$ perturbation of the AD vacua}

 Perhaps the most suggestive result among the vacua of the type (iii) above, is the one discussed in \cite{BGK}.   It was found that the ${\cal N}=2$  Argyres-Douglas  vacua arising from coalescing  $r$ vacua  
 - which is a strongly coupled, nonlocal infrared fixed-point conformal theory of monopoles, dyons and quarks,  when deformed by a simple adjoint mass perturbation, make a metamorphosis into an
 ${\cal N}=1$ 
 superconformal fixed point,   {\it  described by free mesons in the adjoint representation of the flavor $SU(N_f)$ group!  }    Upon further deformation (e.g., some shift of masses),  the system 
 goes into confinement phase,  described by the {\it same} massless degrees of freedom, which now act as the Nambu-Goldstone bosons of the symmetry breaking.

\section{NonAbelian monopoles from monopole-vortex  \\ antimonopole
soliton complex}

\begin{figure}[ht]
\begin{center}
\includegraphics[width=0.7
\linewidth]{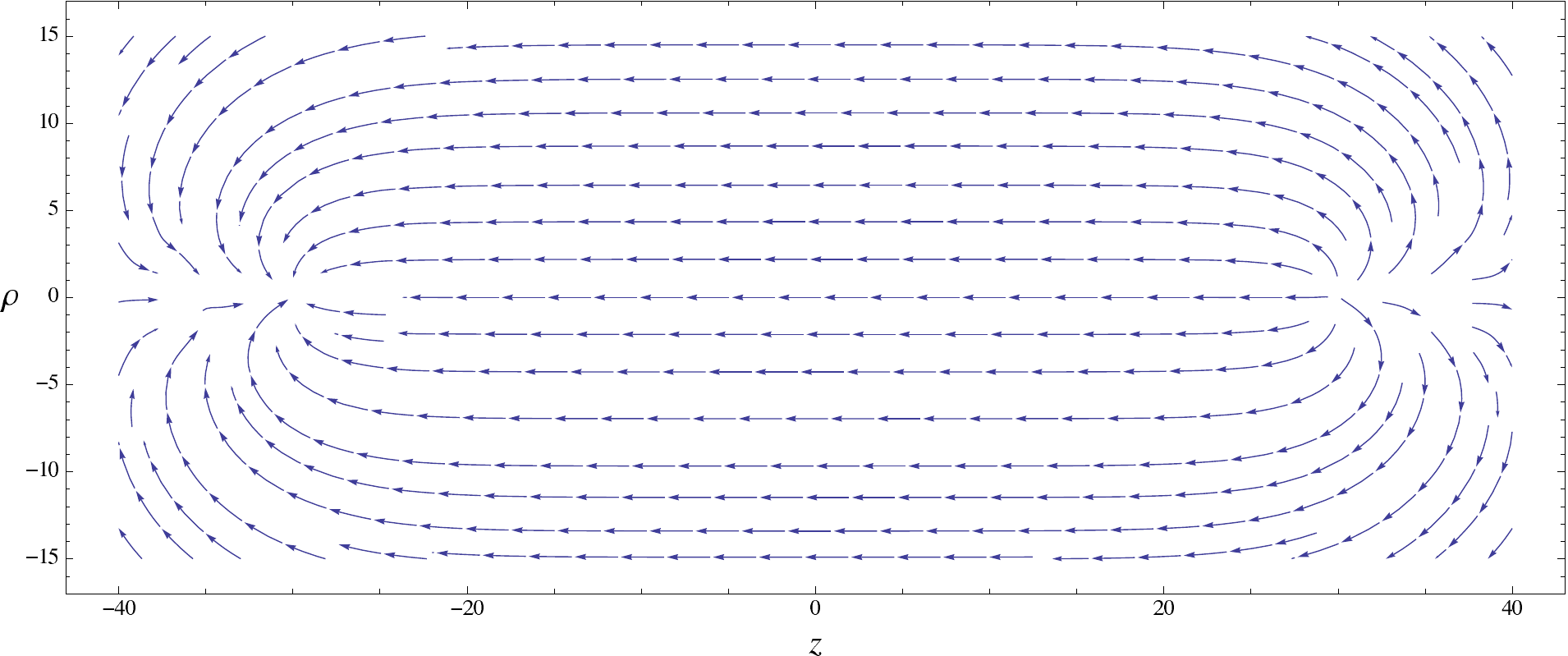}
\caption{\small  The magnetic field in the 
  monopole-vortex-antimonopole soliton complex. Taken from Cipriani, et. al.  \cite{CDGKM}}
\label{MattiaMVA}
\end{center}
\end{figure}

There remains the problem of {\it understanding} the nonAbelian monopoles appearing in the various vacua of ${\cal N}=2$ supersymmetric gauge theories
as the low-energy effective degrees of freedom,  from the
semiclassical point of view, i.e., by relating them to the more familiar picture of topologcal solitons made of the underlying gauge and matter fields.
The key idea  is to consider a hierarchical gauge symmetry breaking, such as 
\be
SU(N+1)_{\rm color} \otimes SU(N)_{\rm flavor} 
  \stackrel{v_{1}}{\longrightarrow} 
(SU(N)\times U(1))_{\rm color}  \otimes SU(N)_{\rm flavor} 
\stackrel{v_{2}}{\longrightarrow} SU(N)_{C+F} \ ,   \label{SUNhierarchy}
\ee
with
\be  
v_{1} \gg v_{2}\ ,  
\ee
as in \cite{ABEKY}-\cite{Konishi}.  The presence of a nontrivial flavor group, and the persistence of an exact global symmetry in a so-called color-flavor locked phase,  is crucial.    A simple homotopy-group consideration shows that the monopoles generated at the high-energy symmetry breaking, having mass of order of   $v_1/ g$,   must be the endpoints of the vortex arising from the low-energy gauge symmetry breaking.  See  Fig.~\ref{MattiaMVA} taken from \cite{CDGKM}.  This must be so as the vortices present in the system at low-energies ($\mu <  v_1$) due to $\pi_1(SU(N)\times U(1))= {\mathbf Z}$
must really be absent in the full theory as  $\pi_1(SU(N+1)={\bf 1}$.  At the same time  the monopoles in the  high-energy  theory (at $\mu \gg v_2$)
due to $\pi_2(SU(N+1)/SU(N)\times U(1))$  are after all absent once the low-energy symmetry breaking is taken into account ($\pi_2(SU(N+1))= {\bf 1}$). 
The homotopy exact sequence connects each monopole to each vortex solution.

The nonAbelian monopoles are to be understood in such a dynamical context, and not just as the properties of static, isolated classic field configurations on their own.  This is perfectly consistent with the fact that the existence of nonAbelian monopoles, being a carrier of a nonAbelian gauge charges, depend critically on the phase of the system.  For instance they would not appear in low energies if the system were to dynamically Abelianize at low energies.

There is actually an even stronger dynamical requirement in order for the concept of nonAbelian monopoles to survive quantum mechanically. 
 This arises from the fact that  the semiclassical  monopoles and vortex,  attached to each other by the homotopy connections, are subject to further quantum fluctuations of the internal, orientational zeromodes, which become important in the far infrared,  $ \mu \ll   v_2$.  
These quantum effects are described by the  $2D$    ${\mathbb C}P^{N-1} $  sigma model,    in the case of the pure 
nonAbelian vortices  \cite{Hanany:2003hp}-\cite{Gorsky:2004ad},   and by the   $2D$-$1D$  coupled   ${\mathbb C}P^{N-1} $  sigma model \cite{CCKK}, in the case of the M-V-M  soliton complex \cite{ABEK}-\cite{CDGKM}
of interest here.  

The monopoles in this context appear in ${\underline N}$ of the  $SU(N)$ isometry group of the  ${\mathbb C}P^{N-1} $.  This fact defines them as nonAbelian monopoles.

The fate of the nonAbelian monopoles as quantum mechanical entity is then linked to the phase of the low-energy  ${\mathbb C}P^{N-1} $  effective action,   defined on a finite-width
worldsheet of the nonAbelian string with a monopole and an antimonopole attached  at its ends.

\section{${\mathbb C}P^{N-1}  $  model in a finite-width worldstrip} 

These considerations have led us recently  to  study \cite{BKO}  the ${\mathbb C}P^{N-1} $  sigma model  defined on the finite-width world strip,   $x \in [0, L]$: 
\be
S= \int dx dt \left( (D_{\mu} n_i)^*D^{\mu} n_i -  \lambda (n_{i}^* n_i - r)   \right) \;,
\ee
where $n^i$ with $i=1,\dots,N$ are $N$ complex scalar fields and the covariant derivative is $D_{\mu} =  \partial_{\mu} -i A_{\mu}$. Configurations related by a  $U(1)$ gauge transformation $z_i \to e^{i\alpha} z_i$ are  equivalent:  the $U(1)$ gauge field $A_{\mu} $ does not have a kinetic term in the classical action.
 $\lambda$ is a  Lagrange multiplier that  enforces the classical condition 
\be
\label{classicalconstrain}
n_{i}^* n_i  = r\;,
\ee
where  $r$ is the ``size'' of the ${\mathbb C}P^{N-1}$ manifold, related to the coupling constant $g$ by  
\be   r= \frac{4 \pi}{g^2}\;. \ee
The quantum properties of this model defined in infinite $2D$  spacetime, with   $x \in [-\infty, \infty]$,  and in the large $N$  approximation  are well known
\cite{D'Adda:1978uc,Witten:1978bc}. 
 Since $n$ appears only quadratically in the Lagrangian, it can be integrated out to give
\beq
Z = \int [dA_{\mu}][d \lambda][d n_i][dn_i^*] \, e^{ i S} =  \int [dA_{\mu}][d \lambda]\,  e^{ i  {S}_{{\rm eff} } }\ ,
\eeq
 leaving an effective action for $A_{\mu}$ and $\lambda$:
\beq
{S}_{{\rm eff}} = \int d^2 x \left( N \, {\rm tr}\, {\rm log} (- D_{\mu}^*D^{\mu}  + \lambda ) -\lambda r\right) \;.
\eeq
The condition of stationarity with respect to $\lambda(x)$ leads to the gap equation (in the Euclidean spacetime) as
\beq
r -N \, {\rm tr} \left(\frac{1}{-\partial_{\tau}^2 -\partial_x^2   +m^2}\right) = 0\;, \label{gap}
\eeq
where we have set $A_{\mu}=0$ and  $\lambda = m^2$. An expectation value of $\lambda$ provides a mass  for the $n_i$ particles.
On the infinite line the spectrum is continuous and the gap equation reads
\beq   \label{large}
r= N  \int_0^{\Lambda_{\rm UV}} \frac{ k dk }{2 \pi } \frac{1}{k^2 +m^2}  =  \frac{N}{4\pi} \log{\left(\frac{\Lambda_{\rm UV}^2 + m^2}{m^2}\right)} \;,
\eeq
 leading to the well-known scale-dependent renormalized coupling
\beq
\label{uvbeta}
r(\mu)= \frac{4 \pi}{g(\mu)^2}    \simeq  \frac{N}{2 \pi}  \log{\left(\frac{\mu}{\Lambda}\right)} 
\eeq
and to a  renormalization-group invariant scale,  $\Lambda$,  in terms of which the dynamically generated mass is given by 
\beq
\label{ml}
\langle \lambda \rangle  =  m^2 = \Lambda^2\;. 
\eeq
Furthermore the kinetic term of the $A_{\mu}$ fields are generated by the one-loop graph, and leads to massless Coulomb field which confines the massive $n$ fields. Thus the spectrum is given by massive $n-n*$ mesons. 

In order to define the model defined on the finite space interval,  $x \in [0, L]$,  one must specify the boundary condition for the $n$ fields.  Motivated by the study of M-V-M complex \cite{CCKK},  we consider the general boundary conditions including  the case of Dirichlet-Dirichlet, 
\beq
\hbox{D-D} : \qquad n_1(0)=n_1(L) = \sqrt{r}\;,   \qquad   n_{i}(0)=  n_{i}(L) =0\;,  \quad i>1\;.  \label{DDbc}
\eeq
and that of  Neumann-Neumann
\beq
\hbox{N-N}: \qquad \partial_x n_i(0) = \partial_x n_i(L) =0\;. 
\eeq
The case of the periodic boundary condition, which has been extensively studied in the literature, is also considered. 

The main question one  asks here  is whether or not the isometry group of the  ${\mathbb C}P^{N-1}$,  $SU(N)$, is spontaneously  broken (Higgs phase) or not  (confinement phase)\footnote{  In the case of DD condition, Milekhin \cite{Milekhin} concluded, by assuming translational invariance Ansatz, that the model has two phases,  the Higgs phase (for $L< L_{crit}$) and a confinement phase (for  $L > L_{crit}$).    
In the case of the model with periodic condition  it was shown \cite{Monin:2015xwa} that the model has two phases,  a Higgs phase at smaller $L< L_{crit}$ and a confinement phase at larger $L > L_{crit}$.     }.

The $N$ fields can be separated into a classical component $\sigma \equiv n_1$ and the rest,   $n^i$    ($i=2,\dots,N$).
Integrating over  the $N-1$ remaining fields  yields  the following effective action
\beq  
{S}_{{\rm eff}} = \int d^2 x \left( (N-1) \, {\rm tr}\, {\rm log} (- D_{\mu}D^{\mu}  + \lambda) + (D_{\mu} \sigma)^*D^{\mu} \sigma -  \lambda (|\sigma|^2  - r)  \right)\;.
\eeq
One can take $\sigma$ real and set the gauge field to zero, and  consider  the leading contribution at large $N$ only.
The total energy  can formally be written as
\be
E =   N   \sum_{n}\omega_n + \int_0^L \left( (\partial_x \sigma)^2 +\lambda (\sigma^2 - r )\right) dx  \;, \label{engen}
\ee  
where $\omega_n^2$ are the eigenvalues of the operator
\beq
\label{operator}
\left( - \partial_{x}^2  + \lambda(x) \right) f_n(x) = \omega_n^2 \ f_n(x)   \;,
\eeq
and $f_n$ are the corresponding   eigenfunctions.    As we are working in the large $N$ approximations we do not distinguish $N-1$ from $N$.  
The eigenfunctions $f_n$ can be taken to be real and orthonormal in  $[0, L]$:
\beq
\int_0^L   dx \, f_n(x) f_m(x) = \delta_{n\,m}\;.
\eeq
The functional variation of $E$ with respect to the fields $\sigma(x)$ and $\lambda(x)$ yields the coupled equations
\beq
\partial_x^2 \sigma(x) - \lambda(x) \sigma(x) = 0, \qquad  \label{gapgen}
\frac{N}{2} \, \sum_n\frac{f_n(x)^2}{\omega_n} + \sigma(x)^2 - r = 0  \,,   
\eeq
which generalizes the gap equation (\ref{gap}).  

The set of equations (\ref{operator})-(\ref{gapgen})  must be solved for two functions  $\lambda(x)$ and $\sigma(x)$.   Apparently this presents a rather formidable  problem:   they represent complicated nonlinear coupled equations. 
We solved these functional saddle-point equations  numerically  by using an idea similar to that of Hartree's approximation in atomic physics.  Namely, starting with some trial $\lambda(x)$, we solve the wave equations (\ref{operator}) for many levels (up to some ultraviolet cutoff mode) and insert them into the second of (\ref{gapgen}).  We find $\sigma(x)$ after subtracting the logarithmic divergence and renormalizing the coupling constant $r$;   the resulting  finite  $\sigma(x)$ is used to obtain $\lambda(x)$ of the next iteration, 
from the first of (\ref{gapgen}). Th procedure can be repeated  until a consistent set of $\lambda(x)$ and $\sigma(x)$ are found. 

A subtle point is that the behavior of the functions $\lambda(x)$ and $\sigma(x)$  near the boundaries $x=0, L$    turns out to be singular, 
\beq
\label{sigmadiv}
 \sigma^2    \simeq   \frac{N}{2 \pi}  \log{\frac{1}{x}}  \;, \qquad  \lambda(x) \simeq    \frac{1}{2  \,x^2 \log{1/x}} \;,  
\eeq
and similarly for $x \to L-x$.  
This result can be traced back to the fact that at the boundaries  where quantum effects are suppressed, $\sigma(x)=n_1(x)$ must account for the 
classical radius of the   ${\mathbb C}P^{N-1}$,  $\sqrt {r}$,  which is large  (see (\ref{DDbc}), (\ref{large})).   The numerical solution described above indeed produces this behavior.

\begin{figure}[!h]
\begin{minipage}{0.42\linewidth}
\includegraphics[width=\linewidth]{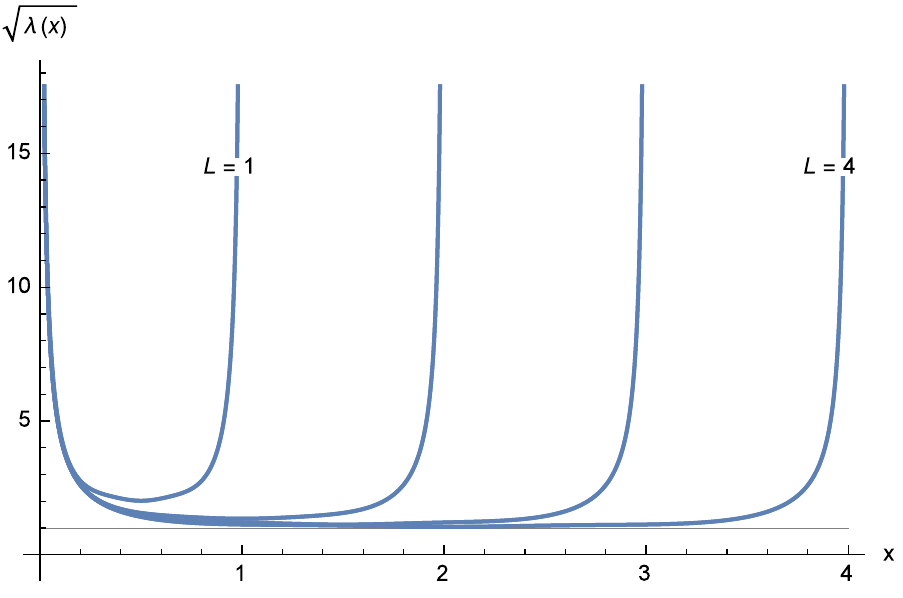}
\caption{The dynamically generated mass of the $n_i$ fields, $\sqrt{\lambda(x)}$. }
\label{lambda}
\end{minipage}
\qquad
\begin{minipage}{0.42\linewidth}
\includegraphics[width=\linewidth]{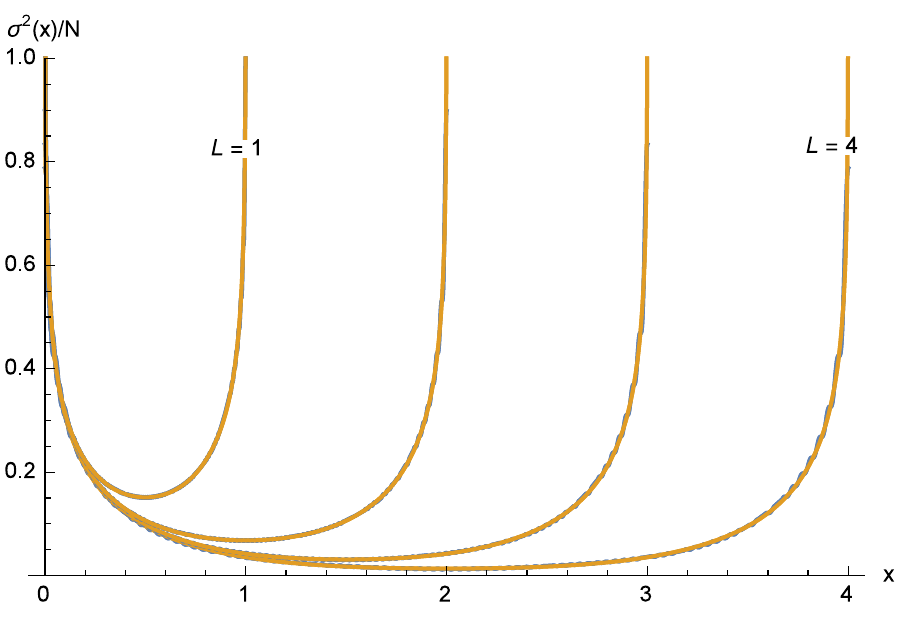}
\caption{The VEV of the $\sigma = n_1$ field.}
\label{sigma}
\end{minipage}
\end{figure}

The results for values of  $ L \Lambda $  up to $4$  are shown  in Figure \ref{lambda}, \ref{sigma},  for $\Lambda=1$ fixed.    From the figures one  sees the expected pattern emerging:  by going to larger $L$ at fixed  $\Lambda$  one expects to recover the confined phase of the standard  ${\mathbb C}P^{N-1}$ sigma model,  Eq.~(\ref{ml}).   We indeed see that $\sqrt{\lambda(x)} \to \Lambda$ in the middle of the interval whereas
the condensate   $\sigma$ approaches zero there at the same time. The effects of the boundary appear to remain concentrated near the two extremes  and not to significantly propagate beyond $1/\Lambda$.

The results found on the ${\mathbb C}P^{N-1} $  sigma model  defined on the finite-width world strip,   $x \in [0, L]$, can be summarized as follows \cite{BKO}. 
\begin{description}
  \item[(i)]   The solution of the generalized gap equations (the functional saddle point equations,  (\ref{operator})-(\ref{gapgen}),  is found to be unique.
  \item[(ii)]    In particular, no "Higgs phase"  solution with   $\langle \lambda \rangle =0\; ,   \langle \sigma \rangle \sim   \Lambda\; \, $ exists. 
  \item[(iii)]  In other words no phase transition from the "confinement phase" with   $\langle \lambda \rangle \sim \Lambda \;,   \langle \sigma \rangle =0\; \, $    to the 
   "Higgs phase" occur.  Our  conclusion differs from that of Milekhin \cite{Milekhin}. The discrepancy may be traced to the fact that 
   due to the boundary condition, the translational invariance Ansatz used in \cite{Milekhin}  is not valid. 
     \item[(iv)]   At large $L$  the  known result  in the standard   ${\mathbb C}P^{N-1} $  sigma model  on infinite $2D$  spacetime,   Eq.~(\ref{ml}),   is seen to emerge from our analysis. 
        \item[(v)]  It turns out that  {\it exactly the same}   results on $\lambda(x)$ and $\sigma(x)$  are found for both DD and NN conditions.  
        Though such a  result may not be obvious at all,  it can be understood by observing that the classical large ${\mathbb C}P^{N-1} $   radius  must be 
        attained independently of the particular form of the boundary condition, as at the boundaries the quantum fluctuations are suppressed. 
        \item[(vi)]   This, and the fact that  the phase transition to the Higgs phase is absent in our case,  can be understood by the fact that the system 
        interpolates between an effectively $1D$  quantum system (quantum mechanics) at  $L \ll  1/\Lambda$  to    genuine $2D$ quantum field theory
          at large $L$,   $L \gg  1/\Lambda$.      
        \item[(vii)]  Dynamical symmetry breaking of the $SU(N)$  isometry group  (the Higgs phase)  does not occur.  This  leaves the nonAbelian nature of the monopoles at the endpoints of the worldstrip  intact:   it transforms as   ${\underline N}$ of $SU(N)$. 
        \item[(viii)]   It was found recently that  this system with {\it periodic boundary condition} allows two possible phases, the Higgs phase at  small  $L$  ($\le  1/ \Lambda$)   and the confinement phase at larger $L$.  We note that even at small $L$  the system maintains $2D$ quantum field theory characteristics in this case, in contrast to the DD or NN conditions discussed above. 
          
\end{description}

\section{Conclusion}

To conclude, the concept of nonAbelian monopole is consistent with quantum mechanics. In a system in which gauge symmetry is broken hierarchically as in (\ref{SUNhierarchy}),   monopoles appear as  ${\underline N}$ of the isometry $SU(N)$  group of  ${\mathbb C}P^{N-1}$,  which describes the effective, low-energy dynamics of the  $M-V-M$ soliton complex.    The $SU(N)$ transformation  corresponds to nonlocal field transformations in terms of the original field variables,  as  the  ${\mathbb C}P^{N-1}$ model describes fluctuations of the collective coordinates. This is perfectly consistent with the electromagnetic duality, albeit in an nonAbelian context.   The isometry  $SU(N)$ may be regarded as a disguise of the dual $SU(N)$ gauge symmetry (in confinement phase)

\section*{Acknowledgment}
The author  thanks the organizers of the Conference, CONF12, 'XII Quark Confinement and the Hadron Spectrum',  Thessaloniki, 
for inviting him to participate and  to give a talk, and for providing such a stimulating atmosphere. The last part of the talk is based on a recent paper in collaboration with  Stefano Bolognesi and Keisuke Ohashi. The rest of the presentation is a summary of the author's earlier (from around 2000) and more recent work  with various collaborators.


\begin{thebibliography}{99}
  
  
  \bibitem{SW1}
N. Seiberg and E. Witten, Nucl.Phys. {\bf B426} (1994) 19; Erratum
\textit{ibid.} \textbf{B430} (1994) 485, hep-th/9407087;  
N. Seiberg and E. Witten, Nucl. Phys. {\bf B431} (1994) 484,
   hep-th/9408099.
   
\bibitem{SUN}
P.~C.~Argyres and A.~F.~Faraggi, Phys. Rev. Lett {\bf 74} (1995)
3931, hep-th/9411047;
A. Klemm, W. Lerche, S. Theisen and S. Yankielowicz, Phys. Lett.
{\bf B344} (1995) 169, hep-th/9411048;
Int. J. Mod. Phys. A11 (1996) 1929-1974, hep-th/9505150;   A. Hanany
and Y. Oz, Nucl. Phys. {\bf B452} (1995) 283,
hep-th/9505075;

\bibitem{APS}   P.C.~Argyres, M.R.~Plesser and N.~Seiberg, Nucl. Phys.  B {\bf 471}, 159  
(1996) [arXiv:hep-th/9603042];  
%P.C.~Argyres, M.R.~Plesser and  A.D.~Shapere, 
%Nucl. Phys. B {\bf 483}, 172 (1997) [arXiv:hep-th/9608129];
 K.~Hori, H.~Ooguri and Y.~Oz,
 Adv. Theor. Math. Phys.   {\bf 1}, 1  (1998) [arXiv:hep-th/9706082].

 
\bibitem{CKM} 
  G.~Carlino, K.~Konishi and H.~Murayama,
  %``Dynamical symmetry breaking in supersymmetric SU(n(c)) and USp(2n(c)) gauge theories,''
  Nucl.\ Phys.\ B {\bf 590}, 37 (2000)
  [hep-th/0005076];
  %%CITATION = HEP-TH/0005076;%%
  %84 citations counted in INSPIRE as of 28 Dec 2013
  G.~Carlino, K.~Konishi, S.~P.~Kumar and H.~Murayama,
  %``Vacuum structure and flavor symmetry breaking in supersymmetric SO(n(c)) gauge theories,''
  Nucl.\ Phys.\ B {\bf 608}, 51 (2001)
  [hep-th/0104064].
  %%CITATION = HEP-TH/0104064;%%
  %37 citations counted in INSPIRE as of 24 Mar 2014
  
\bibitem{SCFT}
 P. C. Argyres,  M. R. Plesser,
N. Seiberg and E. Witten,  {\bf  Nucl. Phys. B  461}   (1996) 71
[arXiv:hep-th/9511154].

% for a recent development, see P.C. Argyres, M. Crescimanno, A.D. Shapere and J.R. Wittig,  ``Classification of N=2 superconformal  field theories with two-dimensional Coulomb branches,''  [arXiv:hep-th/0504070]. 

\bibitem{Eguchi}
 T. Eguchi,  K. Hori, K. Ito and S.-K. Yang, {\bf  Nucl. Phys.  B471}
(1996)
430  [arXiv:hep-th/9603002].

\bibitem{Argyres:2007cn} 
  P.~C.~Argyres and N.~Seiberg,
  %``S-duality in N=2 supersymmetric gauge theories,''
  JHEP {\bf 0712}, 088 (2007)
  [arXiv:0711.0054 [hep-th]].
  %%CITATION = ARXIV:0711.0054;%%
  %100 citations counted in INSPIRE as of 26 Mar 2014 
   
   %\cite{Gaiotto:2010jf}
\bibitem{GST} 
  D.~Gaiotto, N.~Seiberg and Y.~Tachikawa,
  %``Comments on scaling limits of 4d N=2 theories,''
  JHEP {\bf 1101}, 078 (2011)
  [arXiv:1011.4568 [hep-th]].
  %%CITATION = ARXIV:1011.4568;%%
  %29 citations counted in INSPIRE as of 26 Mar 2014
  
  \bibitem{SimoneLorenzo} 
%\cite{DiPietro:2011za}
%\bibitem{DiPietro:2011za} 
  L.~Di Pietro and S.~Giacomelli,
  %``Confining vacua in SQCD, the Konishi anomaly and the Dijkgraaf-Vafa superpotential,''
  JHEP {\bf 1202}, 087 (2012)
  [arXiv:1108.6049 [hep-th]].
  %%CITATION = ARXIV:1108.6049;%%
  %7 citations counted in INSPIRE as of 19 Jun 2014


  \bibitem{SGKK}  
%%\cite{Giacomelli:2013qz}
%\bibitem{Giacomelli:2013qz} 
 S.~Giacomelli and K.~Konishi,
  %``New Confinement Phases from Singular SCFT,''
  JHEP {\bf 1303}, 009 (2013)
  [arXiv:1301.0420 [hep-th]]; 
  %%CITATION = ARXIV:1301.0420;%%
  %2 citations counted in INSPIRE as of 26 Mar 2014%\cite{Giacomelli:2013qz}
%\bibitem{Giacomelli:2013qz} 
%  S.~Giacomelli and K.~Konishi,
%  %``New Confinement Phases from Singular SCFT,''
%  JHEP {\bf 1303}, 009 (2013)
%  [arXiv:1301.0420 [hep-th]].
%  %%CITATION = ARXIV:1301.0420;%%
%  %2 citations counted in INSPIRE as of 26 Mar 2014
% 
  %\cite{Giacomelli:2012ea}
\bibitem{Simone} 
  S.~Giacomelli,
  %``Singular points in N=2 SQCD,''
  JHEP {\bf 1209}, 040 (2012)
  [arXiv:1207.4037 [hep-th]].
  %%CITATION = ARXIV:1207.4037;%%
  %7 citations counted in INSPIRE as of 26 Mar 2014
  

%%\cite{Vainshtein:2000hu}
%\bibitem{Vainshtein:2000hu} 
%  A.~I.~Vainshtein and A.~Yung,
%  %``Type I superconductivity upon monopole condensation in Seiberg-Witten theory,''
%  Nucl.\ Phys.\ B {\bf 614}, 3 (2001)
%  doi:10.1016/S0550-3213(01)00394-7
%  [hep-th/0012250].
%  %%CITATION = doi:10.1016/S0550-3213(01)00394-7;%%
%  %75 citations counted in INSPIRE as of 06 Nov 2016


  
   \bibitem{BGK}
   S.~Bolognesi, S.~Giacomelli and K.~Konishi,
  %``$ \mathcal{N}=2 $ Argyres-Douglas theories, $ \mathcal{N}=1 $ SQCD and Seiberg duality,''
  JHEP {\bf 1508}, 131 (2015)
%  doi:10.1007/JHEP08(2015)131
  [arXiv:1505.05801 [hep-th]].
  %%CITATION = doi:10.1007/JHEP08(2015)131;%%
  %4 citations counted in INSPIRE as of 30 Oct 2016
  
  
  %\cite{'tHooft:1981ht}
\bibitem{tHooft}
  G.~'t Hooft,
  %``Topology Of The Gauge Condition And New Confinement Phases In Nonabelian
  %Gauge Theories,''
  Nucl.\ Phys.\  B {\bf 190}, 455 (1981).:
  %``A Property Of Electric And Magnetic Flux In Nonabelian Gauge Theories,''
  Nucl.\ Phys.\  B {\bf 153}, 141 (1979).: 
  
  \bibitem{CDyons} A. ~Abouelsaood,  Nucl. Phys.  B {\bf 226},  309 (1983);  P. ~Nelson, A. ~Manohar,  Phys. Rev. Lett. {\bf  50},  943
(1983);  A. ~Balachandran, G. ~Marmo, M. ~Mukunda, J. ~Nilsson, E. ~Sudarshan, F. ~Zaccaria,    Phys. Rev. Lett. {\bf  50},  1553
(1983);  P. ~Nelson, S. ~Coleman,  Nucl. Phys.  B {\bf 227},  1  (1984).

%\bibitem{EW}   E. J. Weinberg,   Nucl. Phys.  B {\bf 167}, 500 (1980);  Nucl. Phys.  B {\bf 203}, 445  (1982); 
%K. Lee, E. J. Weinberg, P. Yi,   Phys. Rev. D  {\bf  54 }, 6351  (1996)

\bibitem{DFHK} N. ~Dorey, C. ~Fraser, T.J. ~Hollowood,  M.A.C.~ Kneipp,
% ``NonAbelian duality in N=4 supersymmetric gauge theories,'' 
  Phys.Lett.  B {\bf 383}, 422  (1996)  [arXiv: hep-th/9512116].
  
  \bibitem{GNO}   P. Goddard, J. Nuyts,  D. Olive,    Nucl. Phys. B {\bf  125}, 1 
(1977) 




  \bibitem{Hanany:2003hp}
  A.~Hanany and D.~Tong,
  %``Vortices, instantons and branes,''
  JHEP {\bf 0307} (2003) 037
  [arXiv:hep-th/0306150].

 \bibitem{ABEKY}
R.~Auzzi, S.~Bolognesi, J.~Evslin, K.~Konishi and A.~Yung,
%``Nonabelian superconductors: Vortices and confinement in N = 2 SQCD,''
Nucl.\ Phys.\ B {\bf 673} (2003) 187
[arXiv:hep-th/0307287].
%%CITATION = HEP-TH 0307287;%%

 
 \bibitem{Gorsky:2004ad}
  A.~Gorsky, M.~Shifman, A.~Yung,
%  ``Non-Abelian Meissner effect in Yang-Mills theories at weak coupling,''
  Phys.\ Rev.\  {\bf D71}, 045010 (2005) 
  [arXiv: hep-th/0412082].

    \bibitem{CCKK}
C.~Chatterjee and K.~Konishi,
%  ``Monopole-vortex complex at large distances and nonAbelian duality,''
  JHEP {\bf 1409}, 039 (2014)
%  doi:10.1007/JHEP09(2014)039
  [arXiv:1406.5639 [hep-th]].

\bibitem{ABEK}
  R.~Auzzi, S.~Bolognesi, J.~Evslin and K.~Konishi,
  %``Nonabelian monopoles and the vortices that confine them,''
  Nucl.\ Phys.\  B {\bf 686} (2004) 119
  [arXiv:hep-th/0312233];  
 M.A.C.~Kneipp, 
%  Color superconductivity, Z(N) flux tubes and monopole confinement in deformed N=2* superYang-Mills theories.
Phys. Rev. D {\bf 69}: 045007 (2004) 
[arXiv:hep-th/0308086].

   \bibitem{Duality}
 M.~Eto, L.~Ferretti,  K.~Konishi, G.~Marmorini, M.~Nitta, K.~Ohashi, W.~Vinci and N.~Yokoi,
%  ``Non-abelian duality  from  vortex moduli: 
%a dual model of  color-confinement'',   
Nucl. Phys. B  {\bf 780} 161-187, 2007
   [arXiv:hep-th/0611313].
   
%    \bibitem{CDGKM} 
%  M.~Cipriani, D.~Dorigoni, S.~B.~Gudnason, K.~Konishi and A.~Michelini,
%  %``Non-Abelian monopole-vortex complex,''
%  Phys.\ Rev.\ D {\bf 84}, 045024 (2011)
%  [arXiv:1106.4214 [hep-th]].
   \bibitem{Konishi} K.~Konishi, in Lecture Notes in Physics, {\bf 737}  471 (2008),  Springer  
   [arXiv:hep-th/0702102].     
   

   
  \bibitem{CDGKM}
M.~Cipriani, D.~Dorigoni, S.~B.~Gudnason, K.~Konishi and A.~Michelini,
%  ``Non-Abelian monopole-vortex complex,''
  Phys.\ Rev.\ D {\bf 84}, 045024 (2011)
% doi:10.1103/PhysRevD.84.045024
  [arXiv:1106.4214 [hep-th]];   
  
  
\bibitem{BKO}
%\bibitem{Bolognesi:2016zjp} 
  S.~Bolognesi, K.~Konishi and K.~Ohashi,
  %``Large-$N \mathbb C^{N???1}$ sigma model on a finite interval,''
  JHEP {\bf 1610}, 073 (2016)
%  doi:10.1007/JHEP10(2016)073
  [arXiv:1604.05630 [hep-th]].
  %%CITATION = doi:10.1007/JHEP10(2016)073;%%

\bibitem{D'Adda:1978uc}
  A.~D'Adda, M.~Luscher and P.~Di Vecchia,
%  ``A 1/n Expandable Series of Nonlinear Sigma Models with Instantons,''
  Nucl.\ Phys.\ B {\bf 146} (1978) 63.


\bibitem{Witten:1978bc}
  E.~Witten,
%  ``Instantons, the Quark Model, and the 1/n Expansion,''
  Nucl.\ Phys.\ B {\bf 149} (1979) 285.

  \bibitem{Milekhin} 
  A.~Milekhin,
  %``CP(N-1) model on finite interval in the large N limit,''
  Phys.\ Rev.\ D {\bf 86}, 105002 (2012)
  [arXiv:1207.0417 [hep-th]].
  %%CITATION = ARXIV:1207.0417;%%
  %1 citations counted in INSPIRE as of 29 Apr 2014

% \bibitem{Actor:1985yh}
%  A.~Actor,
%  ``Temperature Dependence Of The Cp**(n-1) Model And The Analogy With Quantum Chromodynamics,''
%  Fortsch.\ Phys.\  {\bf 33} (1985) 333.
%

 \bibitem{Monin:2015xwa}
  S.~Monin, M.~Shifman and A.~Yung,
%  ``Non-Abelian String of a Finite Length,''
  Phys.\ Lev.\ D {\bf 92} (2015) 2,  025011
  %doi:10.1103/PhysLevD.92.025011
  [arXiv:1505.07797 [hep-th]].

  




%%
%\bibitem{GJK} 
%S.~B.~Gudnason, Y.~Jiang and  K.~Konishi,  
%%``Non-Abelian vortex dynamics: Effective world-sheet action'',
% JHEP {\bf 1008}:012 (2010)  
%[arXiv: 1007.2116 [hep-th]].
%
%  
%
%\bibitem{Nambu}
%Y.~Nambu, 
%%lectures at the Copenhagen Summer Symposium,
%%1970 (unpublished); 
%Phys.\ Rev.\ D {\bf 10}, 4262  (1974)
%
%%\bibtem{KalbRam}
%%M.~Kalb and P. Ramond, 
%%Phys.\ Rev.\ D {\bf 9},  2273  (1974)
%
  

  
%\bibitem{Akhmedov}
%%\cite{Akhmedov:1995mw}
%%\bibitem{Akhmedov:1995mw} 
%  E.~T.~Akhmedov, M.~N.~Chernodub, M.~I.~Polikarpov and M.~A.~Zubkov,
%  %``Quantum theory of strings in Abelian Higgs model,''
%  Phys.\ Rev.\ D {\bf 53}, 2087 (1996)
%  [hep-th/9505070].
  %%CITATION = HEP-TH/9505070;%%
  %104 citations counted in INSPIRE as of 18 Apr 2014

  
  
%  \bibitem{Hanany:2004ea}
%  A.~Hanany and D.~Tong,
%  ``Vortex strings and four-dimensional gauge dynamics,''
%  JHEP {\bf 0404}, 066 (2004)
%  [arXiv: hep-th/0403158].
  
  
 
%\cite{Milekhin:2012ca}
%
%%  %
%\bibitem{DKO}  D.~Dorigoni, K.~Konishi and K.~Ohashi,
%%``Non-Abelian vortices with product moduli,''
%   Phys.Rev. D79 (2009) 045011 
%[arXiv: 0801.3284 [hep-th]].

%  
%  \bibitem{Manton}  N.~Manton, Phys. Lett.  B {\bf 154},   397  (1985),   Erratum-ibid. B  {\bf  157}, 475 (1985)
%

      
%  \bibitem{Witten}  
%  E.~Witten, 
%Nucl. Phys. B {\bf 149}, 285  (1979).  
%  
%%\cite{Novikov:1984ac}
%\bibitem{Shifman} 
%  V.~A.~Novikov, M.~A.~Shifman, A.~I.~Vainshtein and V.~I.~Zakharov,
%  %``Two-Dimensional Sigma Models: Modeling Nonperturbative Effects of Quantum Chromodynamics,''
%  Phys.\ Rept.\  {\bf 116}, 103 (1984)
%  [Sov.\ J.\ Part.\ Nucl.\  {\bf 17}, 204 (1986)]
%  [Fiz.\ Elem.\ Chast.\ Atom.\ Yadra {\bf 17}, 472 (1986)].
%  %%CITATION = PRPLC,116,103;%%
%  %141 citations counted in INSPIRE as of 18 Jun 2014  
  
 
   
%  \bibitem{Milekhin:2012ca}
%  A.~Milekhin,
%  ``CP(N-1) model on finite interval in the large N limit,''
%  Phys.\ Lev.\ D {\bf 86} (2012) 105002
%  %doi:10.1103/PhysLevD.86.105002
%  [arXiv:1207.0417 [hep-th]].
 

     
%   \bibitem{MVComplex}
%%\cite{Auzzi:2003em}
%%\bibitem{Auzzi:2003em}
%  R.~Auzzi, S.~Bolognesi, J.~Evslin and K.~Konishi,
%  ``NonAbelian monopoles and the vortices that confine them,''
%  Nucl.\ Phys.\ B {\bf 686} (2004) 119
%%  doi:10.1016/j.nuclphysb.2004.03.003
%  [hep-th/0312233];
%  %%CITATION = doi:10.1016/j.nuclphysb.2004.03.003;%%
%  %84 citations counted in INSPIRE as of 10 Apr 2016
% K.~Konishi, A.~Michelini and K.~Ohashi,
%  ``Monopole-vortex complex in a theta vacuum,''
%  Phys.\ Rev.\ D {\bf 82}, 125028 (2010)
%%  doi:10.1103/PhysRevD.82.125028
%  [arXiv:1009.2042 [hep-th]];  
%  %%CITATION = doi:10.1103/PhysRevD.82.125028;%%
%  %2 citations counted in INSPIRE as of 10 Apr 2016
  
   

\end{thebibliography}
\end{document}